\documentclass[fleqn,usenatbib]{mnras}

\usepackage{newtxtext,newtxmath}

\usepackage[T1]{fontenc}

\DeclareRobustCommand{\VAN}[3]{#2}
\let\VANthebibliography\thebibliography
\def\thebibliography{\DeclareRobustCommand{\VAN}[3]{##3}\VANthebibliography}

\usepackage{graphicx}	
\usepackage{amsmath}	

\usepackage{hyperref}
\usepackage{multirow}
\usepackage{array}
\usepackage{orcidlink}
\usepackage{threeparttablex}

\title[Joint Modelling of X-ray Binary Ejecta]{Joint Radiative and Kinematic Modelling of X-ray Binary Ejecta: Energy Estimate and Reverse Shock Detection}

\author[A. J. Cooper et al.]{A. J. Cooper$^{1 \orcidlink{0000-0002-4033-3139}}$\thanks{E-mail: alexander.cooper@physics.ox.ac.uk},
J. H. Matthews$^{1 \orcidlink{0000-0002-3493-7737}}$, 
F. Carotenuto$^{1,2 \orcidlink{0000-0002-0426-3276}}$, 
R. Fender$^{1,3 \orcidlink{0000-0002-5654-2744}}$, 
G. P. Lamb$^{4 \orcidlink{0000-0001-5169-4143}}$, 
T. D. Russell$^{5 \orcidlink{0000-0002-7930-2276}}$,\newauthor
N. Sarin$^{6,7 \orcidlink{0000-0003-2700-1030}}$, 
K. Savard$^{1 \orcidlink{0009-0001-8598-0639}}$, 
A. A. Zdziarski$^{8 \orcidlink{0000-0002-0333-2452}}$ \\
$^{1}$Astrophysics, The University of Oxford, Keble Road, Oxford, OX1 3RH, UK\\
$^{2}$INAF-Osservatorio Astronomico di Roma, Via Frascati 33, I-00078, Monte Porzio Catone (RM), Italy\\
$^{3}$Department of Astronomy, University of Cape Town, Private Bag X3, Rondebosch 7701, South Africa \\
$^{4}$Astrophysics Research Institute, Liverpool John Moores University, IC2 Liverpool Science Park, 146 Brownlow Hill, Liverpool, L3 5RF, UK \\
$^{5}$INAF, Istituto di Astrofisica Spaziale e Fisica Cosmica, Via U. La Malfa 153, I-90146 Palermo, Italy \\
$^{6}$Oskar Klein Centre, Department of Physics, Stockholm University,
Albanova University Center, SE 106 91 Stockholm, Sweden \\
$^{7}$Nordita, Stockholm University and KTH Royal Institute of Technology, Hannes Alfvéns väg 12, SE-106 91 Stockholm, Sweden\\
$^{8}$Nicolaus Copernicus Astronomical Center, Polish Academy of Sciences, Bartycka 18, PL-00-716 Warszawa, Poland \\
}

\date{Accepted XXX. Received YYY; in original form ZZZ}

\pubyear{\the\year{}}

\begin{document}
\label{firstpage}
\pagerange{\pageref{firstpage}--\pageref{lastpage}}
\maketitle

\begin{abstract}
Black hole X-ray binaries in outburst launch discrete, large-scale jet ejections which can propagate to parsec scales. The kinematics of these ejecta appear to be well described by relativistic blast wave models original devised for gamma-ray burst afterglows. In previous kinematic-only modelling, a crucial degeneracy prevented the initial ejecta energy and the interstellar medium density from being accurately determined. In this work, we present the first joint Bayesian modelling of the radiation and kinematics of a large-scale jet ejection from the X-ray binary MAXI J1535-571. We demonstrate that a reverse shock powers the bright, early ejecta emission. The joint model breaks the energetic degeneracy, and we find the ejecta has an initial energy of $E_{0} \sim 3 \times 10^{43} \, {\rm erg}$, and propagates into a low density interstellar medium of $n_{\rm ism} \sim 4 \times 10^{-5} \, {\rm cm^{-3}}$. The ejecta is consistent with being launched perpendicular to the disc and could be powered by an efficient conversion of available accretion power alone. This work lays the foundation for future parameter estimation studies using all available data of X-ray binary jet ejecta. 
\end{abstract}

\begin{keywords}
ISM: jets and outflows -- X-rays: binaries -- shock waves -- acceleration of particles -- gamma-ray burst: general -- radio continuum: transients
\end{keywords}



\section{Introduction}
Astrophysical jets are observed ubiquitously from accreting compact objects across a variety of spatial scales. Supermassive black holes (BHs) at the centres of galaxies power Mpc-scale jets \citep[e.g.,][]{blandford_2019} which evolve on Myr timescales and whose feedback appears to regulate galactic growth \citep[e.g.,][]{Ferrarse_2000}. Stellar-mass BHs can accrete rapidly after their formation to power gamma-ray bursts (GRBs), or more slowly from companion stars through winds or Roche lobe overflow as BH X-ray binaries (XRBs). 
\par
BH-XRBs are binary systems in which stellar-mass BHs ($M_{\rm BH} \lesssim 20 M_{\odot}$) accrete from a companion star. BH-XRBs are typically identified by wide-field X-ray instruments during outbursts likely triggered by disc instabilities (e.g., \citealt{lasota_2001}). During outburst, BH-XRBs cycle through X-ray states on timescales of days-months \citep{remillard_2006}. Canonically (see e.g., \citealt{fender_2004}), these sources begin in quiescence before entering the hard state upon outburst onset, characterised by hard spectrum X-ray emission ($L_{\rm X} \lesssim 0.1 L_{\rm Edd}$) and a partially absorbed compact, steady radio jet with a flat spectral index. The X-ray luminosity increases ($L_{\rm X} \gtrsim 0.1 L_{\rm Edd}$), before the X-ray spectrum begins to soften and the source enters a bright intermediate phase. The softening of the X-ray emission marked by dramatic changes in the jet \citep{fender_2009}, where the compact jet is quenched, bright self-absorbed radio flares are produced, and large-scale relativistic ejecta are launched along the jet axis. These ejecta are observed as discrete knots of optically thin, synchrotron emitting `blobs' of plasma. The source will usually enter the soft state during which no compact jet emission is detected, and the X-ray luminosity begins to decrease. The source transitions back to the hard state as the X-ray luminosity usually continues to fade towards quiescence. During this reverse soft-to-hard transition, observations are consistent with no further ejecta being launched, instead the compact jet is gradually re-established upon return to the hard state \citep[e.g.,][]{russell_2014}. The production and propagation of the relativistic jet ejecta during the hard-soft transition are the focus of this work.  
\par
Bipolar jet ejections from the BH-XRB GRS 1915+105 were identified by \cite{mirabel_1994} as the first apparently superluminal Galactic sources. The advent of large-scale GHz radio interferometers has, through dedicated programmes such as ThunderKAT \citep{fender_2016}, enabled the regular detection and tracking of discrete ejecta as they propagate out to large distances from the core. Sixteen such sources now have discrete ejecta resolved from the core at either radio or X-ray frequencies \citep{mirabel_1994, hjellming_2000,hannikainen_2001, corbel_2002, mioduszewski_2001, gallo_2004,corbel_2005,yang_2010,rushton_2017,russell_2019,miller-jones_2019, bright_2020, carotenuto_2021,williams_2022, wood_2023, bahramian_2023,zhang_2025}, with a subset tracked continuously to core-offsets of tens of arcseconds. However, theoretical interpretation and modelling of the excellent available data has been somewhat limited. The first model of BH-XRB ejecta was pioneered by \cite{wang_2003} via adaptation of the kinematic GRB model from \cite{huang_1999} to fit the eastern jet of XTE J1550$-$564. That model seeks to conserve the conical jet's energy as it sweeps up mass and decelerates:
\begin{equation}
    E_0 = (\Gamma - 1) M_0 c^2 + \sigma (\Gamma_{\rm sh} - 1) m_{\rm sw} c^2
\end{equation}
where $E_0$ and $M_0$ are the jet's initial energy and mass respectively, $m_{\rm sw}(t)$ is the swept-up mass, $\sigma$ is the adiabatic index (interpolated between ultra-relativistic and non-relativistic values), and $\Gamma(t)$ and $\Gamma_{\rm sh}(t)$ are the Lorentz factors of the jet material and shock front respectively. \citet{wang_2003} inferred that a low interstellar medium (ISM) density (also known as the circumburst density) is required to explain the observed propagation of the ejecta from XTE J1550$-$564, indicative of parsec-scale under-dense cavities in the environment surrounding the source. The authors further modelled the X-ray emission of the ejecta, followed by similar modelling of both the approaching and receding jets by \cite{hao_2009}. In their work, \citet{wang_2003} find that a forward shock model evolved too slowly to explain the observed X-ray flux, but a reverse shock fit the data relatively well. 
\par
Since these seminal works, modelling attempts have exclusively focused on the kinematics of XRB jets. \citet{steiner_2012} used a conical jet model to further study the jet kinematics of XTE J1550$-$564, allowing them determine a jet inclination angle ($\theta_{\rm obs}$) of $\sim 71$ degrees. The same methodology was used by \citet{steiner_2012b} to determine a distance and inclination angle for double-sided jets of H1743$-$322. \citet{carotenuto_2022a} recently adapted this model to include a sharp density jump representing the edge of the under-dense cavity surrounding MAXI J1348$-$630 in their modelling of jet kinematics, although \citet{zdziarski_2023} showed that a smooth cavity transition explains the observations while alleviating energetic requirements. \citet{carotenuto_2024} applied the same kinematic model to three more XRB sources with large-scale ejecta: MAXI J1820+070, MAXI J1535$-$571, and XTE J1752$-$223. The authors were able to determine jet launching times, initial Lorentz factors, and upper limits to the ISM density, $n_{\rm ism}$, by comparing degenerate energy estimates to available accretion power over launching timescales inferred by radio flare durations. However, these kinematic-only, conical jet models have a key degeneracy between the initial energy $E_0$, the half-opening angle $\theta_{\rm c}$, and the density of the ISM. This means that usually only a degenerate quantity known as the `effective energy' $E_{\rm 0, eff} = E_0 n_{\rm ism}^{-1} (\theta_{\rm c})^{-2}$ can be derived \citep{carotenuto_2024}, unless one or more of these quantities is measured by other means. Finally, \cite{Sarath_2025} utilised a kinematic model similar to \citet{zdziarski_2023} to model the ejecta from MAXI J1348$-$630, while employing a modified blazar model to predict the radiative properties. Their model successfully explains the late-time radio rebrightening of the ejecta upon ejecta-cavity interaction, but does not capture the early-time flux which we attribute in this work to a reverse shock.
\par
Interpretating the full wealth of excellent data from BH-XRB jets is crucial to better understand their nature. The supreme data quality, owing to proximity and evolution on human-accessible timescales probes the propagation and deceleration in detail enabling a unique view of macrophysical blastwave physics and the microphysics of particle acceleration in jets \citep{matthews_2025}. Furthermore, excellent prior observational constraints, particularly of jet opening angles, distances, and BH properties greatly enhance modelling capabilities. 
This is particularly important in order to verify long standing notions that we may be greatly underestimating the true energetics of BH-XRB jets \citep{gallo_2005}, and that BH-XRBs are surrounded by under-dense cavities \citep{heintz_underdense_2002,hao_2009,carotenuto_2022a,carotenuto_2024,savard_2025}. More accurate determinations of the initial jet energies will also help quantify their importance as a Galactic feedback mechanism \citep{heinz_2002,heinz_2007,heinz_2008}, and as a significant source of high-energy cosmic rays and neutrinos \citep{fender_2005,cooper_2020,kimura_2021,kantzas_2023,kuze_2025,bacon_2025}. The latter is of particular relevance given recent detections of ultra high-energy ($>100$ TeV) gamma-rays from a number of BH-XRBs \citep{LHAASO:2024psv}, in addition to earlier detections of very high-energy ($>100$ GeV) gamma-rays \citep{hess_2005,hess_2006,albert_2007,Abeysekara_2018,ss433_tev_2018}. Finally, a better understanding of Galactic BH jets may enable us to better characterize other jetted sources including GRBs, tidal disruption events, and active galactic nuclei, where similar (often scale-invariant) physics dictates jet propagation and particle acceleration.
\par
In this work, we present the first combined radiative and kinematic modelling with Bayesian parameter estimation through a case study of jet ejecta observed from a BH-XRB, MAXI J1535$-$571. In Section \ref{sect:sources} we discuss the MAXI J1535$-$571 system. In Section \ref{sect:model_data} we discuss the model, the data selection, and the fitting procedure, including our choice of prior. In Section \ref{sect:results} we present our results, which we discuss in Section \ref{sect:discussion}. We present our primary conclusions and future outlook in Section \ref{sect:conclusion}. In Appendix \ref{app:inclination} we include further discussion on the jet inclination angle of MAXI J1535-571, and in Appendix \ref{app:plots} we include additional diagnostic plots.

\section{MAXI J1535-571}
\label{sect:sources}

MAXI J1535$-$571 (henceforth MAXI J1535) is a BH-XRB, discovered after going into outburst in September 2017 \citep{negoro_2017,nakahira_2018,Tao_2018} by the Monitor of All-sky X-ray Image (MAXI; \citealt{matsuoka_2009}) and the SWIFT Burst Alert Telescope (Swift/BAT; \citealt{gehrels_2004}). The outburst was followed across the electromagnetic spectrum in X-ray \citep{huang_2018,nakahira_2018,miller_2018,parikh_2019,Sreehari_2019,dong_2022}, optical/infrared \citep{dincer_2017,russell_2017,vincentelli_2021}, and radio/sub-mm \citep{russell_2017,russell_2019,russell_2020,chauhan_2021}. 
\par
We chose to apply this new joint modelling technique to MAXI J1535 primarily due to the wealth of data on the transient ejecta and the observational constraints available on the source. \citet{russell_2019,russell_2020} presented a dedicated radio campaign using ATCA (Australia Telescope Compact Array) and MeerKAT (`More' Karoo Array Telescope), tracking the discrete jet ejection (labeled S2 in their work) for over 300 days. The high spatial resolution of the instrumentation enabled the authors to resolve the core-ejecta separation providing ejecta flux measurements that were uncontaminated from the core emission. 
\par
The authors fit the proper motion to constrain the launch date of the ejecta to between MJD 58001.7 and MJD 58026.7 depending on whether the ejecta motion is uniform or decelerating. \citet{carotenuto_2024} fit a conical blastwave model to the kinematic data presented in \citet{russell_2019}, constraining the ejecta launch date to MJD $58017.4^{+4.0}_{-3.8}$, the initial Lorentz factor to $\Gamma_0 = 1.6^{+0.2}_{-0.2}$, and the effective energy to  $E_{\rm 0, eff} = 5.8^{+16.6}_{-4.0} \times 10^{48}\, (n_{\rm ism}/{\rm 1 \, cm^{-3}}) \, (\theta_{\rm c}/1 \, {\rm deg})^{-2} \, {\rm erg}$. We note that this ejection date overlaps with the peak in the X-ray lightcurve \citep{shang_2019} and the possible detection of Type-B quasi-periodic oscillations (QPOs) reported by \citet{stevens_2018} from stacked NICER data across MJD 58016.8-58025. Type-B QPOs are thought to be possibly associated with transient jet launching (see e.g., \citealt{Motta_2016,ingram_motta_19,wood_2021}) and support an ejection date at least in the latter third of the 15 day window reported by \citet{russell_2019}, consistent with the launch date derived by \citet{carotenuto_2024}. \citet{chauhan_2021} presented quasi-simultaneous broadband radio observations spanning $117\,$MHz -- $19\,$GHz from MJD 58016--58040, spanning the jet ejecta launch and early evolution. The authors interpret their observations as flaring events associated with the launching of discrete ejecta, and as possible fading emission from the ejecta. Improved the spatial resolution of observations over this time may distinguish between these scenarios although we note that the emission fades by MJD 58030, so while it is probably associated with the same material which goes on to produce the large-scale ejecta, it is unlikely in-situ particle acceleration has begun. Finally, \cite{chauhan_2019} carried out observations of HI absorption spectrum, determining a best-fit distance to the source of $4.1^{+0.6}_{-0.5}$ kpc, with a robust upper limit at $6.7^{+0.1}_{-0.2}$ kpc, and a lower limit of $3.6\,$kpc.

\subsection{On the inclination angle of the disc \& jets of MAXI J1535}
\label{sect:inclination_angle_prediscussion}
X-ray measurements of disc reflection features were carried out on data obtained by various facilities using \texttt{XSPEC} \citep{arnaud_1996}. NICER observations of the production region of narrow Fe K emission line using \texttt{relline} model imply an inclination angle of $\theta_{\rm obs} = 37^{+22}_{-13}$ degrees, yet in the same work fits of an absorbed disk blackbody component with a relativistically blurred reflection using the \texttt{relxill} model find an angle of $\theta_{\rm obs} = 67.4(8)$ degrees \citep{miller_2018}. NuSTAR observations found two models provided satisfactory fits to the data (\texttt{xillverCp} and \texttt{relxillCp}) with derived inclination angles of $\theta_{\rm obs} = 57^{+2}_{-1}$ degrees and $\theta_{\rm obs} = 75^{+4}_{-2}$ degrees respectively \citep{xu_2018}. Finally, a joint NuSTAR and HXMT analysis by \cite{dong_2022} derived a best fit inclination angle of the inner accretion disc of $\sim 70-74$ deg across all datasets. Broadly, these results  typically required large inclination angles $\theta_{\rm obs} > 45$ degrees. 
\par
\cite{russell_2019} present constraints on the viewing angle of the transient ejecta, finding $\theta_{\rm obs} < 45$ degrees. The authors use the maximum measured proper motion of $47.2$ mas day$^{-1}$, and solve for families of solutions in $\theta_{\rm obs}$ and $\beta$ for various distances measurements compatible with results of \citet{chauhan_2019}. They use the standard equation for superluminal motion \citep{rees_1966,mirabel_1994}:
\begin{equation}
    \label{eq:rel_beta_theta}
    \mu_{\rm app} = \frac{\beta \sin(\theta_{\rm obs})}{1 - \beta \cos(\theta_{\rm obs})} \frac{c}{D}
\end{equation}
Solutions compatible with the observed proper motion are depicted in their figure 9, and appear in tension with X-ray disc inclination measurements. This led the authors to posit disc warping, or a possible disc-jet misalignment in MAXI J1535. In our re-analysis, we derived an extension to the acceptable family of solutions above $\theta_{\rm obs} = 45$ deg which can reproduce the required proper motion. This re-analysis considerably weaken previous constraints, to $\theta_{\rm obs} < 76$ degrees for a distance of $4.7$ kpc, but unconstrained for the closest distance considered of $3.6$ kpc. For this reason, we consider a prior for $\theta_{\rm obs}$ encompassing the full 0-90 degrees. Further details are provided in the Appendix \ref{app:inclination}. 

\section{Joint Radiative and Kinematic Model}
\label{sect:model_data}

\subsection{Ejecta model}
To model both the radiation and kinematics of the discrete ejecta we utilise \texttt{jetsimpy}\footnote{\href{https://jetsimpy.readthedocs.io/en/latest/}{https://jetsimpy.readthedocs.io}} \citep{wang_2024}. \texttt{jetsimpy} is an efficient, reduced hydrodynamic code which approximates the blast wave as an axisymmetric, infinitely thin, two-dimensional surface. Primarily designed for modelling GRB afterglows, \texttt{jetsimpy} produces lightcurves and kinematic profiles for trans-relativistic outflows, calibrated to self-similar, ultra-relativistic \citep{blandford_mckee_1976} and Newtonian \citep{sedov_1959} blast-wave solutions, making it suitable for XRB ejecta. For the kinematic modelling, we compute at each time $t$ the angular separation of the ejecta from the core using the \texttt{jet.Offset} function. We note that while this separation only solves for the forward shock, at early times when emission is dominated by the reverse shock, the forward and reverse shocks are likely co-located (see also subsection \ref{sect:caveats} and \citealt{matthews_2025}).
\par
For the radiation modelling, we compute the emission from the forward shock utilising \texttt{jetsimpy}'s deep-Newtonian synchrotron \texttt{sync\_dnp} radiation model \citep{sironi_2013}. This choice is made over the default \texttt{sync} model, as it has been shown to more faithfully reproduce late-time flattening of GRBs \citep{sironi_2013} and to provide better fits to late-time lightcurve of the GW170817 afterglow \citep{ryan_2024}. These deep Newtonian corrections, which modify the temporal decay of radio flux, are partcularly important to include for XRBs due to their generally lower $\Gamma$ blast waves as compared to GRBs.

\par
For the reverse shock emission, we construct a custom radiation model within \texttt{jetsimpy}. We adopt the thin reverse shock model of \cite{kobayashi_2000a} (see also 3.1.2. of \citealt{gao_2013}), as it is more appropriate than thick shock models for lower Lorentz factors, expected for mildly relativistic XRB ejecta. We additionally correct the canonical (ultra-relativistic) reverse shock crossing timescale to account for moderate initial Lorentz factors (equation 17 in \citealt{matthews_2025}). We compute first the expected on-axis flux in this regime, taking into account all relevant hierarchies of critical frequencies, where $\nu_{\rm m}$ is the synchrotron minimum frequency, $\nu_{\rm a}$ is the synchrotron self-absorption frequency, and $\nu_{\rm c}$ is the cooling frequency. We transform the on-axis observer-frame flux into the co-moving ejecta frame to calculate a deboosted flux ($F_{\nu, {\rm co-moving}}(t) \approx F_{\nu, { \theta_{\rm obs} = 0}}(t)/(2 \Gamma(t))^3$), before dividing through by the forward shock volume as calculated by \texttt{jetsimpy} to obtain an emissivity. A crucial assumption here is that the forward and reverse shock widths are equivalent ($\delta R_{\rm FS} \sim \delta R_{\rm RS}$). We examine this assumption in Model (A6) in which we fit for a `filling factor' parameter $f = \delta R_{\rm RS}/\delta R_{\rm FS}$. The observer-frame flux density is calculated from this emissivity which takes into account the observer viewing angle, photon time-of-arrival effects, and optional jet-spreading (lateral jet expansion to regions $\theta > \theta_{\rm c}$). Note that we omit order unity numerical correction coefficients derived by \cite{harrison_13}. This may lead to a slight overestimate of the reverse shock flux, however, given the ejecta's low bulk $\Gamma$, this correction is expected to be nominal.

\subsection{Data selection}
\label{sect:data_selection}
We fit the model to the radio data of the jet ejection `S2' presented by \citet{russell_2019}. We use all 12 core separation datapoints in their table 5 for the kinematic modelling. Errors quoted for separation datapoints are statistical only, thus for simplicity we assign additional conservative $5\%$ errors to each to approximate the systematic error\footnote{Future works, particularly where the ratio of separation to flux datapoints is larger or kinematic fits are unsatisfactory, may approximate systematic error using a percentage of the beam size where available.}. For the flux measurements, we use all 67 datapoints including non-detections of the ejecta. For the uncertainties, we use the 1$\sigma$ errors included for detections, and non-detections are three times the root-mean square (rms) of the image noise. We do not include early-time observations (MJD 58000--58050) reported by \citet{chauhan_2021} as these data, taken prior to the detection of distinct ejecta component S2, may be contaminated by emission from the core. Furthermore, these early-time observations are likely probing synchrotron emission from particles accelerated during the initial flare or in a \citet{vanderlaan_66} type phase, whereas our model captures the in-situ particle acceleration powered by either the forward or reverse shock.

\subsection{Fitting framework}
We utilise nested sampling \citep{Skilling_2004} package \texttt{dynesty}\footnote{\href{https://dynesty.readthedocs.io/en/stable/}{https://dynesty.readthedocs.io/en/stable/}} \citep{speagle_2020} to ensure full exploration of posterior parameter space and to enable model comparison. We employ a standard Gaussian likelihood function:
\begin{equation}
    {\rm ln} [L(\theta, \xi)] = -\frac{n}{2} {\rm log}(\sigma^2) - \frac{1}{2} \sum_{i=1}^{n} \bigg(\frac{x_i - \mu}{\sigma}\bigg)^2
\end{equation}
Where $x_{n}$ are the model fluxes and separations (which we jointly fit), $\mu$ and $\sigma$ are the data and errors respectively. For non-detections, we fit for a flux of $0$ with errors consistent with the observed upper limits. We use $1024$ live points, random walk sampling (\texttt{rwalk}), and the default multi-ellipsoidal decomposition (\texttt{multi}) which were found to provide a good balance between flexibility and computational cost. All fits were run with a consistent stopping criterion (\texttt{dlogz = $0.001$}), where \texttt{dlogz} is the log of the ratio between the current estimated evidence and the remaining evidence. For all models, we present the natural logarithm of the Bayesian evidence (marginal likelihood) $\ln(Z)$ and the $\chi^2/{\rm dof}$ values, where the degrees of freedom are the number of datapoints (79) less the number of free parameters.  

\subsection{Priors}
In Table \ref{tab:priors} we detail the prior constraints used for modelling. For the initial bulk Lorentz factor $\Gamma_0$ we use a log-uniform prior to ensure the model does not bias towards larger Lorentz factors unless required by the data. For the inclination angle we adopt an isotropic (cosine\footnote{We note possible confusion between cosine/sin priors. This prior assumes isotropy which maximizes probabilities of jets at $\theta_{\rm obs} = 90$ deg and drops to zero at $\theta_{\rm obs} = 0$ deg.}) prior, as expected from geometric arguments for an unknown value of $\theta_{\rm obs}$ (see discussion in Sections \ref{sect:inclination_angle_prediscussion} and Appendix \ref{app:inclination}). We do not include priors based on X-ray fits to disc emission/reflection due to the inconsistency between values derived from different models and the possibility of jet/disc misalignment. For the source distance we adopt a prior based on the derived distance by \cite{chauhan_2019}, with a normal distribution truncated at the authors' derived minimum and maximum distances, opting for a minimum distance of $3.5$kpc to account for the possibility of an association with a SNR located at this distance \citep{maxted_2020}. We opt to initially fit for magnetic energy fractions in the forward and reverse shocks individually, as GRB fits imply these values can differ \citep{2003_zhang,harrison_13,huang_2016,2019_lamb_a,2019_lamb_b}. These magnetic energy fractions $\epsilon_{\rm B}$ are chosen to have a large log-uniform range with a maximum at values corresponding to equipartition values e.g., $\epsilon_{\rm B, eq} = 0.5$ \footnote{For a charge neutral electron-proton jet the maximal equipartition value for $\epsilon_{\rm B}$ and $\epsilon_{\rm e}$ would be $1/3$, nevertheless we use $0.5$ as a formal upper limit to conservatively account for the possibility of non-neutral or leptonic jets.}. For the surrounding ISM density we adopt a wide log-uniform prior to encompass very low densities predicted and inferred in cavities around X-ray binary jets \citep[e.g.,][]{hao_2009,carotenuto_2024,savard_2025}. We have three fixed parameters in the initial models, where the first two (the jet half-opening angle $\theta_{\rm c}$ and the fractional energy transferred to electrons $\epsilon_{\rm e}$) are fixed to avoid degeneracies in posterior distributions. Around the launching time \cite{chauhan_2021} were able to resolve the ejecta source size through long baseline radio observations and, by including observed proper motion, confidently identify the opening angle of the ejecta (independent of the angle to the line of sight). We fix this parameter to their best-fit value of $4.5$ deg. The fraction of energy in electrons is fixed to the maximum allowed value in equipartition of $0.5$. We choose to adopt this maximum value for the initial models, such that our derived ejecta energy $E_{0}$ can be considered a minimum value: the energy in non-thermal particles is set by $\epsilon_{\rm e}$ and the blastwave energy which depends on $E_{\rm 0}$. This means that generally, lower values of $\epsilon_{\rm e}$ will result in a commensurately higher value of $E_{\rm 0}$\footnote{Fixing the $\epsilon_{\rm e}$ value has implications for the evolution of critical synchrotron frequencies. However, these frequencies are stable over the jet evolution (see Appendix \ref{app:criticalfreqs}), meaning the result of variations in $\epsilon_{\rm e}$ for these models is limited to the normalization of the flux. We verify this by testing additional models in which $\epsilon_{\rm e}$ is allowed to vary freely (see Section \ref{sect:further_investigation}).}. Finally, the launch time of the ejecta is fixed based on kinematic-only constraints of \citet{russell_2019} and \citet{carotenuto_2024}, and the radio/X-ray flaring activity observed at the time\footnote{We note that the degeneracy in $\theta_{\rm obs}$ and $\beta$ solution means that despite \cite{carotenuto_2024} finding a best fit value of $\theta_{\rm obs} < 45 \deg$, fixing the ejecta launch time to this date remains a reasonable choice (see Appendix \ref{app:inclination} for more details).}.

\setlength{\tabcolsep}{8pt}
\setlength{\extrarowheight}{.7em}
\begin{table*}
\caption{List of parameters for the initial models (A-D) with prior distributions and bounds. For later runs (A2-A6), the same free parameters and priors are used as here, unless explicitly stated in the text.}
\label{tab:priors}


\begin{tabular}{*{4}{c l c c}}

\hline
\hline
\textbf{Parameter}  & \textbf{Description} 	&   \textbf{Prior} & \textbf{Bounds/Value}\\
\hline                                  
$\Gamma_0$ & Initial bulk Lorentz factor & Log-uniform & [1,10]\\
${E}_{\rm iso, min}$ & Isotropic-Equivalent Initial Energy [erg] & Log-uniform & [$10^{40}$,\,$10^{50}$] \\
$\theta_{\rm obs}$ & Jet Inclination Angle (degrees) & Cosine & [$0$,$90$] \\
$D$     & Source distance (kpc) &  Normal [Truncated] & $4.1_{-0.6}^{+0.6}$ [$3.5$,\,$6.7$]\\
$n_{\rm ism}$ & ISM (Circumburst) Density [${\rm cm}^{-3}$]& Log-uniform & [$10^{-8}$, $1$]\\
$\epsilon_{\rm B, FS}$ & Fraction of energy in forward shock magnetic field & Log-uniform & [$10^{-8}$,\,$0.5$]\\
$\epsilon_{\rm B, RS}$ & Fraction of energy in reverse shock magnetic field & Log-uniform & [$10^{-8}$,\,$0.5$]\\
$p$ & Power-law of accelerated electrons & Uniform & [$2.2$,\,$4.2$] \\
$\theta_{\rm c}$ & Half-opening angle of jet [degrees] & \textbf{Fixed} & \textbf{2.25} \\
$\epsilon_{\rm e}$ & Fraction of energy in accelerated electrons & \textbf{Fixed} & \textbf{0.5}\\
$t_{\rm ej}$ & Launch time of ejecta & \textbf{Fixed} & \textbf{MJD 58017.4}\\
\hline         
\end{tabular}


\end{table*}

\section{Results}
\label{sect:results}

\subsection{Initial modelling}
We initially fit four models, comprised of combinations of two jet profile morphologies (tophat \& Gaussian) corresponding to uniform and Gaussian distributed jet energy (as defined by equations 59 \& 60 in \citealt{wang_2024}), and turning lateral jet spreading on or off. These models are identical with the exception of jet morphology and spreading choices, and thus can be compared directly. Full details of these different jet models are available in \citet{wang_2024}. In Table \ref{tab:fitting_results_1} we present the marginal likelihood, $\log(Z)$, and reduced Chi-squared, $\chi^2/{\rm dof}$, values for best-fit parameter set obtained by each model. In each case, we also compute the Bayes Factor (ratio between marginal likelihoods) as compared to the best-performing Model (A). A Bayes factor of $\mathcal{B}$ can be interpreted as the null model (A) being a factor of $\mathcal{B}$ more likely as compared to the test model. Values of $\mathcal{B}$ greater than 100 imply decisive support for Model (A), the null hypothesis \citep{kass_bayes_1995}. We find conclusive evidence in favour of jets where spreading is turned off, as measured by both the reduced $\chi^2$ and large Bayes factor values ($\mathcal{B}_{\rm BA} \gg 1$). The Bayes factor obtained comparing Model (C) to Model (A), $\mathcal{B}_{\rm CA} = 223$, implies weaker, yet still significant evidence that the data is best explained by a tophat rather than Gaussian profile jet.
\par
\begin{figure*}
\begin{center}
\includegraphics[width=\textwidth]{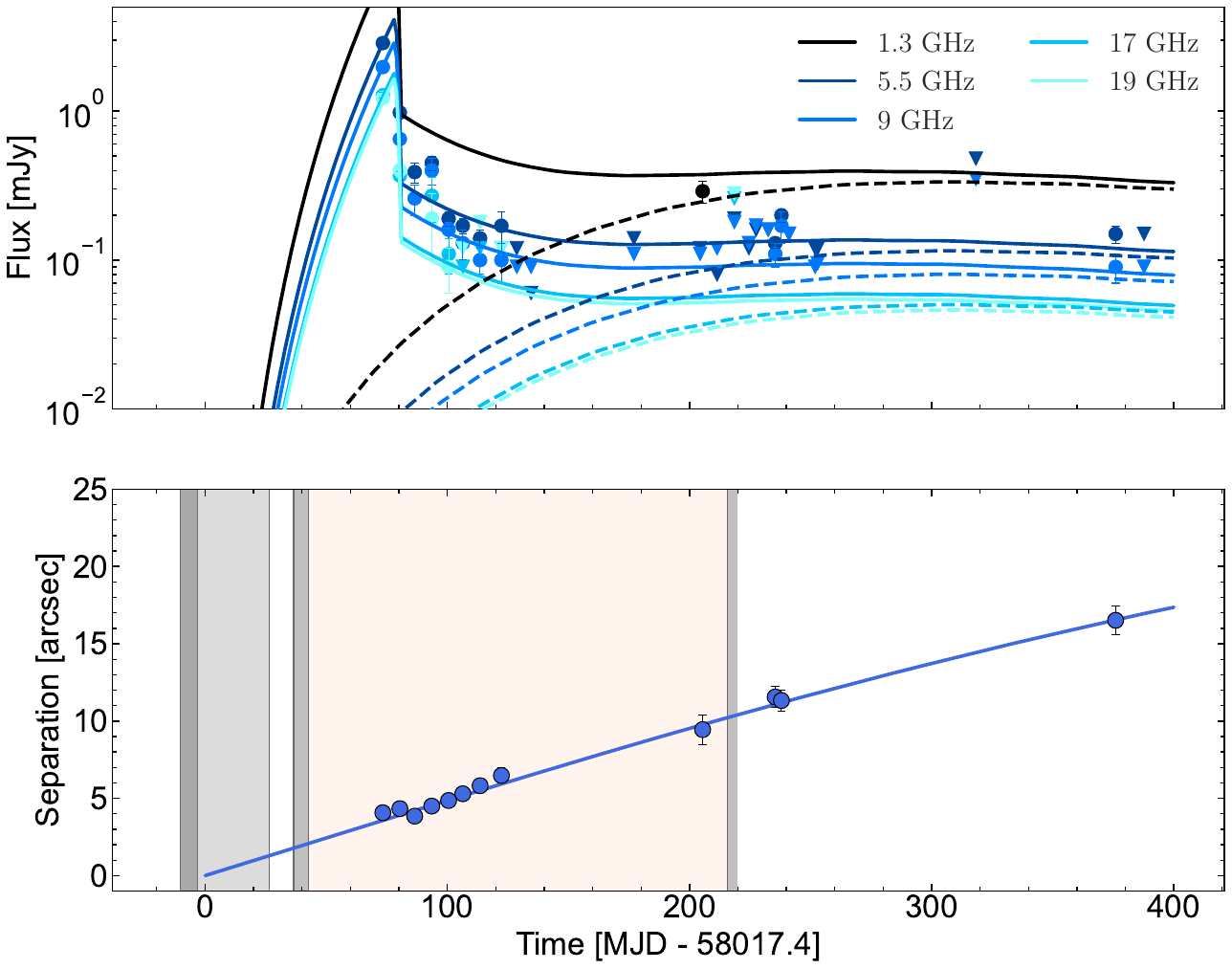}
\caption{Preferred Model (A) lightcurves (top panel) and core-separation (bottom panel). In the top panel the total flux density (reverse shock + forward shock) at observed frequencies is shown by the solid lines (lowest frequency = darkest, highest frequency = lightest), with the forward shock component denoted by dashed lines. Background shades in the bottom panel correspond to the X-ray state of MAXI J1535 from \protect\cite{Tao_2018}: unshaded, dark grey, light grey, and seashell shades correspond to the hard, hard-intermediate, soft-intermediate, and soft states respectively.}
\label{fig:lc_best_fit}
\end{center}
\end{figure*}

In Fig. \ref{fig:lc_best_fit} we present the best-fit Model (A) lightcurve and kinematic fits, where in the top panel the total (forward + reverse shock) flux density is denoted by solid lines (coloured by frequency) and the forward shock component is highlighted as dashed lines. The reverse shock dominates initial emission before 100 days, with a sharp rise during the reverse shock crossing. After the reverse shock crosses, no new particles are accelerated and the flux declines sharply and $\nu_{\rm m}$ decreases. Shortly thereafter it decreases below $\nu_{\rm a}$ resulting in the emission fading more gradually, before the forward shock begins dominating emission. 
Both the reverse and forward shock emission are optically thin and slow cooling throughout (e.g., the hierarchy $\nu_{\rm a} < \nu_{\rm obs} < \nu_{\rm c}$ holds throughout for both the reverse and forward shock at radio frequencies), with the exception of a short optically thick phase early in the evolution. This phase is very short (less than 10 days) for the preferred Model (A) and slightly longer for other models, but always shorter than the reverse shock crossing duration. This consistent hierarchy is expected based on the relatively stable observed spectral index of the jet ejecta \citep{russell_2019}, and means the evolution of lightcurves in all models (except the forward shock only Model A2) is driven primarily by the reverse shock crossing, the blastwave dynamical evolution, and the late-time brightening of the forward shock. Further discussion on the temporal evolution of the critical frequencies can be found in Appendix \ref{app:criticalfreqs}. All four models tested follow similar profiles and best-fit lightcurves and separation are presented in Appendix \ref{app:plots}.
\par
In Table \ref{tab:fitting_results_best_fit_params} we present the best fit values for the free parameters, quoting 1$\sigma$ uncertainties assuming a Gaussian posterior distribution. Here we briefly discuss the parameters, leaving most of the interpretation to Section \ref{sect:discussion}. In all models, a relatively low initial Lorentz factor is derived, consistent with the lower range obtained by \citet{carotenuto_2024}, with spreading jets requiring slightly higher values of $\Gamma_0$ and $E_{0}$ as expected. Consistently low ISM densities are derived and relatively high energies with minimum true (non-isotropic; $E_{\rm true} = E_{\rm iso} \theta_{\rm c}^2/4$) initial ejecta energies ranging from $E_{\rm 0, min} = 3.4 - 5.9 \times 10^{43} \, {\rm erg}$. Taking the effective energy derived by kinematic-only modelling \citep{carotenuto_2024} and substituting our derived Model (A) values of $n_{\rm ism}$ and fixed value of the half-opening angle, we find $E_{0, \rm C24} = 4.89^{+14.0}_{-3.37} \times 10^{43}\,{\rm erg}$, consistent with initial energies derived in this work. Our Model (A) initial energy and Lorentz factor imply a total initial ejecta mass of $M_0 = E_0 / (\Gamma_0 - 1) c^2 = 1.1 \times 10^{23} \, {\rm g} \approx 5.5 \times 10^{-11} \, M_{\odot}$. 
\par
In all models the derived inclination angles are $\theta_{\rm obs} \sim 70$ deg, in agreement with the X-ray disc observations. The distance measurements are consistently pushed towards lowest value allowed in the prior of $3.5$ kpc, which is also observed in the posterior distributions (see discussion in Section \ref{sect:caveats}). Finally, the values of the electron powerlaw $p$ are relatively consistently placed around a reasonable value of $2.55$, consistent with the observed spectral index of $\alpha \sim -0.75$ (where $F_{\nu} \propto \nu^{\alpha}$ and $p = 1 - 2\alpha$ for $\nu_{\rm obs} > \nu_{\rm m}$) in \citet{russell_2019}. Surprisingly, the reverse shock magnetization, $\epsilon_{\rm B, RS}$, is very small, whereas typically in GRBs $\epsilon_{\rm B, RS} > \epsilon_{\rm B, FS}$. We run additional models in Section \ref{sect:further_investigation} to test this finding, and further discussion is continued in Section \ref{sect:discussion}.


\setlength{\tabcolsep}{8pt}
\setlength{\extrarowheight}{.9em}
\begin{table*}
\centering
\begin{tabular}{l|l|l|l|l}
\hline
\hline
\textbf{Jet Model} & \textbf{Sub-type} & \textbf{log(Z)} & \textbf{Reduced $\chi^2$} & \textbf{Bayes Factor} \\ \hline 
Tophat & No Spreading (A) & $\textbf{24.68} \pm \textbf{0.25}$ & \textbf{1.57} & \textbf{1} (null) \\ \cline{2-5} 
                  & Spreading (B) & $-10.89 \pm 0.25$ &  2.59 & $\sim 10^{15} $  \\ \hline
Gaussian & No Spreading (C) & $19.27 \pm  0.26$ & 1.64 & $223$ \\ \cline{2-5} 
                    & Spreading (D) & $-0.75 \pm 0.26$  &  2.20 & $\sim 10^{11}$  \\ \hline
\end{tabular}
\caption{Results of the initial modelling of four jet profiles, where tophat profile, non-spreading jet (Model A) performs best both in terms of the goodness of fit and Bayes factor.}
\label{tab:fitting_results_1}
\end{table*}

\setlength{\tabcolsep}{3pt}
\setlength{\extrarowheight}{0.9em}
\begin{table*}
\centering
\begin{tabular}{l|l|l|l|l|l|l|l|l}
\hline
\hline
\textbf{Jet Model}  & ${\rm log}_{10} E_{\rm 0, min}$ & $\Gamma_0$ & ${\rm log}_{10} n_{\rm ism}$ &  $\theta_{\rm obs}$ [deg] & ${\rm log}_{10} \epsilon_{\rm B, RS}$ & $p$ & ${\rm log}_{10} \epsilon_{\rm B, FS}$ & D [kpc] \\ \hline 

\textbf{Tophat No Spread} (A) & $43.54^{+0.58}_{-0.49}$ & $1.38^{+0.02}_{-0.02}$ & $-4.40^{+0.58}_{-0.50}$ & $71.89^{+2.20}_{-2.66}$ & $-6.06^{+1.07}_{-1.24}$ & $2.48^{+0.06}_{-0.05}$ & $-1.87^{+1.07}_{-1.24}$ & $3.56^{+0.06}_{-0.03}$ \\

Tophat Spread (B) & $43.75^{+0.53}_{-0.53}$ & $1.44^{+0.02}_{-0.02}$ & $-4.34^{+0.54}_{-0.54}$ & $79.41^{+1.98}_{-1.85}$ & $-6.25^{+1.14}_{-1.16}$ & $2.56^{+0.06}_{-0.06}$ & $-1.95^{+1.14}_{-1.15}$ & $3.52^{+0.02}_{-0.01}$\\ 

Gaussian No Spread (C) & $43.53^{+0.59}_{-0.49}$ & $1.41^{+0.02}_{-0.02}$ & $-4.51^{+0.60}_{-0.48}$ & $65.62^{+2.23}_{-2.32}$ & $-5.82^{+1.03}_{-1.30}$ & $2.53^{+0.05}_{-0.05}$ & $-1.75^{+1.04}_{-1.28}$ & $3.54^{+0.04}_{-0.02}$ \\ 

Gaussian Spread (D) & $43.77^{+0.60}_{-0.56}$ & $1.46^{+0.02}_{-0.02}$ & $-4.42^{+0.61}_{-0.56}$ & $73.87^{+2.00}_{-2.12}$ & $-6.13^{+1.19}_{-1.29}$ & $2.60^{+0.05}_{-0.05}$ & $-1.96^{+1.17}_{-1.30}$ & $3.52^{+0.02}_{-0.01}$ \\ \hline
\end{tabular}
\caption{Parameter estimation for the initial models (A-D). We quote the median of the posterior distribution, with errors consistent with the 1$\sigma$ confidence interval. Note that the minimum initial energy is the true (beaming-corrected) energy.}
\label{tab:fitting_results_best_fit_params}
\end{table*}

\begin{figure*}
\begin{center}
\includegraphics[width=\textwidth]{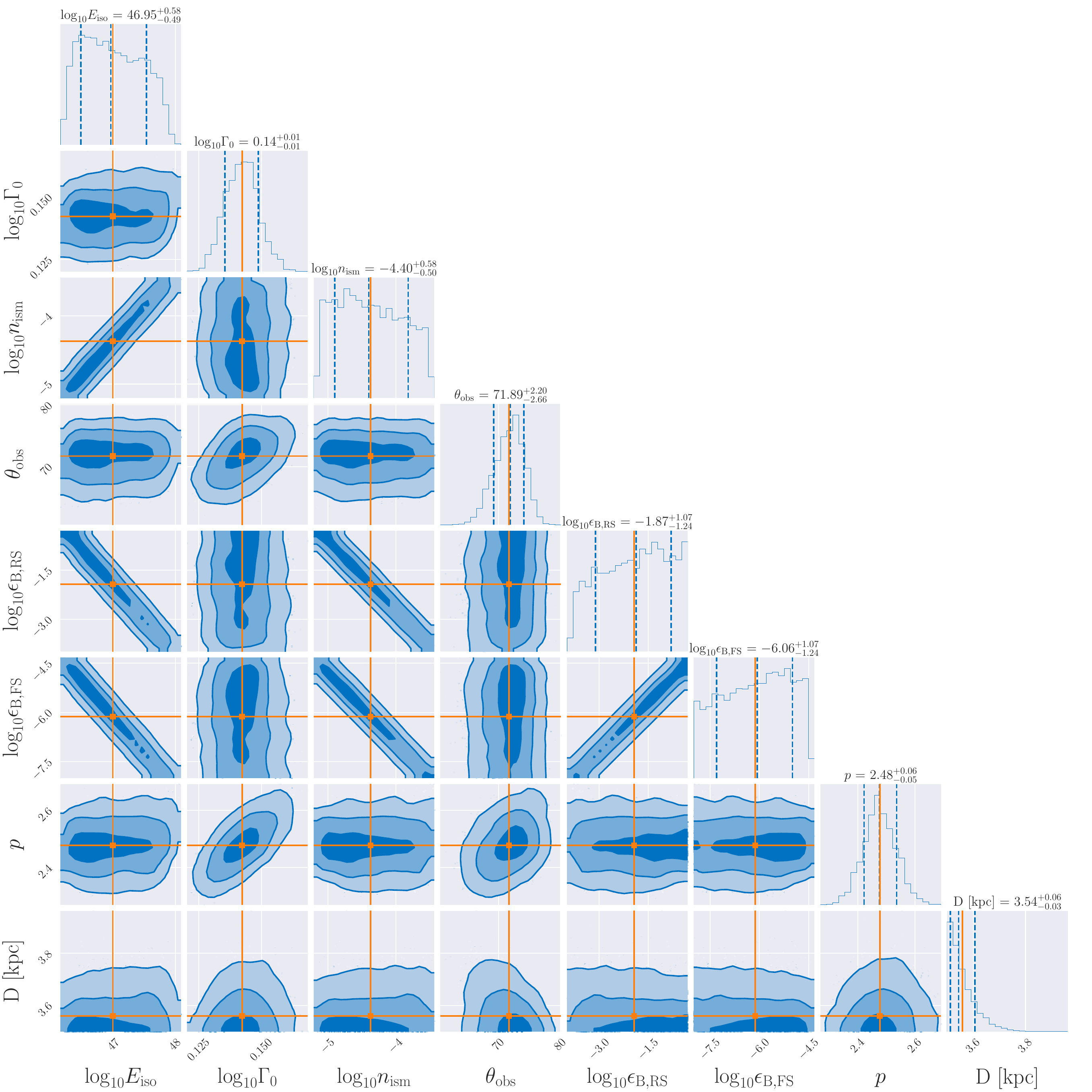}
\caption{Corner plot of the posterior distributions for the best-fit Model (A). The best-fit value (mean of the posterior samples) is denoted in orange, with the median and 1$\sigma$ confidence interval (e.g., containing 68\% of posterior probability mass) with blue dashed lines. Only limited degeneracies are seen in key energetic parameters including $E_{\rm iso}$, $n_{\rm ism}$ and $\epsilon_{\rm B}$. The best-fit distance is consistent with the minimum values of the prior constraint of $3.5$ kpc.}
\label{fig:corner_tophat_spread}
\end{center}
\end{figure*}

\subsection{Further investigation}
\label{sect:further_investigation}
To interrogate our results, we test additional models with a variety of free parameter set-ups, using the best-fit Model (A) scenario of a tophat jet without spreading. The results of these fits are shown in Table \ref{tab:fitting_results_2}, with lightcurves and separations in Appendix \ref{app:plots}. We compute the reduced $\chi^2$ and Bayes factor compared to Model (A) as before, but as model variations have varying numbers of free parameters, we also compute the Bayesian Information Criterion (BIC; \citealt{schwarz1978}):
\begin{equation}
    {\rm BIC} = \kappa ln(n) - 2 ln(L(\hat{\theta}))
\end{equation}
where $\kappa$ is the number of free parameters, $n=79$ is the number of data points, and $ln(L(\hat{\theta}))$ is the log-likelihood of the parameters $\hat{\theta}$ that maximize the likelihood (e.g. best-fit parameters). Lower relative values of BIC indicate preferred models and, unlike the Bayes factor, the BIC does not depend on the prior distribution.

\subsubsection{Single shock models}
In the first two (A2 \& A3), we run Model (A) with only the forward shock (A2) and reverse shock (A3) contributing to emission. Both models perform poorly and are heavily disfavoured as compared to the combined Model (A), with the reverse shock only model performing better due to its ability to capture all early time data. Modification of our assumption of a uniform density ISM (e.g., without a cavity wall) may significantly improve the fit obtained by the forward shock only model, albeit with additional free parameters. Nonetheless, we argue the reverse shock is necessary for the following reasons. Firstly, the forward shock only Model (A2) requires unrealistic parameters, including $\Gamma_{0} \gtrsim 10$ and $p \gtrsim 4.2$ (both posterior distributions bunch at the maximum prior value) where $p$ especially is inconsistent with the observed spectral index. Crucially, the $p$ value is pushed to unphysically high values merely to capture the steep decline in flux density, rather than matching the observed spectral index. Secondly, the forward shock only model predicts a very bright peak in emission prior to the first detection (see Fig. \ref{fig:all_best_fit_curves}). However, the core flux detected at this time had a flat spectral index and occurred during a return to the X-ray hard state. This suggests the emission is consistent with compact jet emission, rather than early ejecta emission. Moreover, one of these significant core detections occurs on MJD 58059 when the telescope is in a high spatial resolution configuration `6A', which should have resolved the ejecta. Finally, the model performs very poorly compared to reverse shock models even in explaining only early-time lightcurve and separation, implying multiple changes in density profile would be required to explain the data, including rebrightenings. Further discussion, including of the impact of a cavity wall on the reverse shock only model, is continued in Section \ref{sect:caveats}.

\subsubsection{Free global $\epsilon_{\rm e}$}
In Model (A4), we include $\epsilon_{\rm e}$ as an additional free parameter, and enforce a wider prior for $\epsilon_{\rm e}$ as well as $\epsilon_{\rm B, FS}$ \& $\epsilon_{\rm B, RS}$ to allow all values up to 1, no longer enforcing an equipartition maximum. This model performs well, yet the base model (A) with fixed $\epsilon_{\rm e}$ is still marginally preferred by all metrics. This, in addition to the consistently high uncertainties in parameter estimation, implies that the data is not sufficient to support the inclusion of additional free parameters. In other words, a (limited) degeneracy is present between $E_{0}$, $\epsilon_{\rm e}$, and $n_{\rm ism}$ as hypothesized. Interestingly, despite magnetization parameters remaining consistent with Model (A), a lower value of $\epsilon_{\rm e} \approx 0.1$, albeit with large uncertainties, is preferred. This change is commensurate with an increase in the initial energy ($E_{0, \rm A4} \approx 8 \times 10^{43}\, {\rm erg}$), suggesting the initial energy should be slightly higher than our lower limit of $E_{0, \rm A} \approx 4 \times 10^{43}\, {\rm erg}$ found in Model (A). 

\subsubsection{Global $\epsilon_{\rm B}$, free $\epsilon_{\rm e}$}
In Model (A5), we instead fit for a global $\epsilon_{\rm B}$ which is applied to both the forward and reverse shock, and allow a freely varying $\epsilon_{\rm e}$ parameter for each shock. This model is run to test the hypothesis whether the low derived value of $\epsilon_{\rm B, RS}$ obtained in Model (A) merely reflects a lower particle acceleration or radiative efficiency (e.g., whether lower value of $\epsilon_{\rm e, RS}$ can also reproduce the data). The model performs well, with the BIC and reduced $\chi^2$ values very similar to Model (A), although the Bayes factor is disfavoured. The derived value of $\epsilon_{\rm e, RS} \sim 10^{-2.7}$ and higher values of $\epsilon_{\rm e, FS} \sim 10^{-0.9}$ and $\epsilon_{\rm B} \sim 10^{-0.2}$ suggest we cannot draw any robust conclusions regarding the low inferred $\epsilon_{\rm B, RS}$ from Model (A). In other words, we cannot distinguish the exact mechanism which limits the efficiency of the reverse shock emission with the present data, but merely that it is possibly due to low $\epsilon_{\rm e}$, $\epsilon_{\rm B}$, or both. The corner plot for this model is included in Appendix \ref{app:plots}.

\subsubsection{Free reverse shock width}
Finally, in Model (A6), we aim to test our underlying assumption that the shock thickness $\delta R_{\rm FS} \sim \delta R_{\rm RS}$. To do this, we again fix $\epsilon_{\rm e}$ and fit for different $\epsilon_{\rm B}$ values for each shock, but include a freely varying filling factor parameter $f = \delta R_{\rm RS}/\delta R_{\rm FS}$. This parameter is allowed to vary log-uniformly between [$10^{-6}, 1$], where a value of 1 corresponds to the forward shock width computed by \texttt{jetsimpy}, e.g. Model (A). Variations in $\delta R_{\rm RS}$ in the model effectively result in commensurate changes the emissivity (as this calculated by dividing the rest-frame flux by the reverse shock volume) of the reverse shock without otherwise affecting the radiation calculations, unlike changes to $\epsilon_{\rm e}$. Across all metrics, Model (A6) performs well, but Model (A) is still marginally preferred. 

The estimated filling factor is around $\delta R_{\rm RS}/\delta R_{\rm FS} \sim 0.19^{+0.45}_{-0.13}$ suggesting the base model is not vastly underestimating the volume of reverse shocked material, validating our methodology.

\setlength{\tabcolsep}{8pt}
\setlength{\extrarowheight}{.9em}
\begin{table*}
\centering
\begin{tabular}{l|l|l|l|l|l}
\hline
\hline
\textbf{Tophat No Spread} & \shortstack{ \\ \textbf{Free} \\ \textbf{Params}} & \textbf{log(Z)}  & \textbf{Reduced }$\chi^2$ & \textbf{BIC} & \textbf{Bayes Factor} \\[2ex] \hline 
(A) FS + RS Base model & 8 & \textbf{24.68} $\pm$ \textbf{0.25} & $1.57$ & \textbf{-77.96} & \textbf{1} (null) \\ \hline
(A2) FS Only & 7  & $-890.43 \pm 0.22$ & 27.13  & 1759.30& $\sim 10^{400}$  \\ \hline
(A3) RS Only & 7 & $-69.55 \pm 0.23$ & 4.29 & 115.01 & $\sim 10^{40}$ \\ \hline
(A4) FS + RS (free $\epsilon_{\rm e}$\protect\footnotemark) & 9 &  $22.80\pm0.26$ & 1.63 & $-71.09$ & $6.5$ \\ \hline
(A5) FS + RS (free $\epsilon_{\rm e, RS}$, $\epsilon_{\rm e, FS}$, $\epsilon_{\rm B}$) & $9$ & $16.81 \pm 0.28$ & \textbf{1.56} & $-76.12$ & 2618 \\ \hline
(A6) FS + RS (free $\delta R_{\rm RS}/\delta R_{\rm FS}$)& 9 & $24.43 \pm 0.25$ & 1.58  & -74.57 & 1.28 \\ \hline
\end{tabular}
\caption{Results of the modelling of additional variations on the best-fit Model (A), the details of which can be found in the discussion in Section \ref{sect:further_investigation}.}
\label{tab:fitting_results_2}
\end{table*}
\footnotetext{We relax the prior constraint in models where both $\epsilon_{\rm e}$ and $\epsilon_{\rm B}$ are fit to allow all values $\epsilon \leq 1$.}

\setlength{\tabcolsep}{3pt}
\renewcommand{\arraystretch}{1.4}
\begin{table*}
\centering
\begin{tabular}{l|l|l|l|l|l|l|l}
\hline
\hline
\textbf{Jet Model}  & ${\rm log}_{10} E_{\rm min}$ & ${\rm log}_{10} n_{\rm ism}$ & ${\rm log}_{10} \epsilon_{\rm B, (RS)}$ & ${\rm log}_{10} \epsilon_{\rm B, FS}$ & ${\rm log}_{10} \epsilon_{\rm e, (RS)}$ & ${\rm log}_{10} \epsilon_{\rm e, FS}$ & $\delta R_{\rm RS}/\delta R_{\rm FS}$\\ \hline 

Model (A) &  $43.54^{+0.58}_{-0.49}$ & $-4.40^{+0.58}_{-0.50}$ & $-6.06^{+1.07}_{-1.24}$ & $-1.87^{+1.07}_{-1.24}$ & 0.5 (Fixed) & n/a & n/a \\ \hline 

Model (A4) free $\epsilon_{\rm e}$ & $43.90^{+0.46}_{-0.65}$ & $-3.98^{+0.48}_{-0.66}$ & $-5.55^{+1.02}_{-1.48}$ & $-1.36^{+1.01}_{-1.49}$ & $-0.92^{+0.63}_{-0.71}$ & n/a & n/a \\ \hline  

Model (A5) free $\epsilon_{\rm e, RS}$, $\epsilon_{\rm e, FS}$, $\epsilon_{\rm B}$ & $43.10^{+0.31}_{-0.27}$ & $-4.87^{+0.30}_{-0.26}$ & $-0.20^{+0.14}_{-0.20}$ & n/a & $-2.65^{+0.15}_{-0.20}$ & $-0.88^{+0.61}_{-0.84}$ & n/a  \\ \hline 

Model (A6) free $\delta R_{\rm RS}/\delta R_{\rm FS}$ & $43.42^{+0.53}_{-0.37}$ & $-4.59^{+0.55}_{-0.37}$ & $-6.98^{+1.17}_{-0.73}$ & $-1.48^{+0.80}_{-1.15}$ & $0.5$ (Fixed) & n/a & $-0.71^{+0.52}_{-0.95}$  \\  
\hline
\end{tabular}
\caption{Parameter estimation for Models (A) \& (A4-A6) for energetically relevant parameters. Quoted values are the median of the posterior distribution, with errors consistent with the 1$\sigma$ confidence interval. Note some models fit for a global $\epsilon_{\rm B/e}$, whereas some fit independently for the forward and reverse shocks.}
\label{tab:fitting_results_best_fit_params_further}
\end{table*}

\section{Discussion}
\label{sect:discussion}
\subsection{Initial energy and jet-launching mechanism}
Our best-fit initial energy, from Model (A), is $E_{0, \rm min} = 3.48^{+9.7}_{-2.36} \times 10^{43}$ erg, increasing to $E_{0} = 7.90^{+15.01}_{-6.13} \times 10^{43}$ erg for Model (A4) where $\epsilon_{\rm e}$ is allowed to vary freely, which may better reflect the true initial ejecta energy. We can compare these energy (and mass) estimates to the accretion flow properties over the jet-launching timescale to attempt to discriminate between disc powered (\citealt{blandfordp_82}; BP82) and BH spin powered jets (\citealt{blandfordz_77}; BZ77). 
\par
Self-absorbed radio flares are often observed around the launch of large-scale ejecta \citep{fender_2009,fender_bright_19}, and their rise duration may correspond to the approximate launching timescale of the ejecta such that $t_{\rm launch} \leq t_{\rm flare, rise}$. A bright radio flare was observed from MAXI J1535 around the launch of the ejecta component on MJD 58017. In addition, \cite{chauhan_2021} reported bright radio emission beginning to rise at 58017.17 and \cite{russell_2019} reported fading radio emission during a 3.5hr observation starting at MJD 58017.38. This implies a maximum flare rise time of approximately $t_{\rm flare, rise} \lesssim 5$ hours and thus an ejection launch time less than $5$ hours. This is in agreement with discrete ejecta launching timescales derived for MAXI J1820+070 through observations of quenching X-ray variability, Type B QPOs \citep{homan_2020} and very long baseline imaging \citep{wood_2021}.
\par
In the following we compare the mass and energy budget of the ejecta to the disc properties across the launching timescale. The estimated mass of the ejecta in our best-fit Model (A) is $1.1\times 10^{23}$ g. The kinematically inferred launch time corresponds to a peak in the Swift/XRT X-ray lightcurve reported by \cite{shang_2019} on MJD 58017.03. The authors fit a two component advective flow (TCAF; \citealt{tcaf_1995}) model and, assuming a fitted BH mass of $M_{\rm BH} = 8.1 M_{\odot}$, derive a mass accretion rate of $\dot{M}_{\rm acc} \approx 6.5 \times 10^{18} \; {\rm g \, s^{-1}}$ where $\dot{M}_{\rm acc} = L_{\rm acc}/c^{2}$. We can compare this to MAXI observations presented by \cite{sridhar_2019}, in which the authors reported $F_{\rm bol}=1.6\times 10^{-7}$ erg cm$^{-2}$ s$^{-1}$ on MJD 58007. Their figure 1 shows that the MAXI flux doubled from MJD 58007-58017; therefore we estimate $F_{\rm bol}\approx 3\times 10^{-7}$ erg cm$^{-2}$ s$^{-1}$ during the jet launching. Assuming $D\approx4$ kpc, we estimate $L_{\rm bol}\approx 6\times 10^{38}$ erg s$^{-1}$, and $\dot M_{\rm accr}\approx 6.4\times 10^{18} \; {\rm g \, s^{-1}}$, approximately the same as \cite{shang_2019} for an accretion efficiency of 0.1.
\par
Assuming the ejection duration of $\delta t_{\rm launch}=5$ hours, the mass accreted is $1.2\times 10^{23}$ g, only slightly larger than the ejecta mass. Under the assumption that the mass is accreted onto the BH (e.g., is not entrained in the jet), the total accretion required is $\dot M_{\rm tot} = \dot{M}_{\rm ejecta} + \dot{M}_{\rm acc} \approx 1.3 \times 10^{19} \; {\rm g \, s^{-1}}$. This requires an extremely efficient (50\%) mechanism of ejecting wind from the disk. We can compare this requirement with the $\dot{M}_{\rm wind}$ predicted by equation 5.2 of \citet{blandfordp_82}. We assume that the magnetic field is in a MAD (magnetically arrested disc) state, given by the magnetic flux, $2\pi B R_{\rm g}^2$ and  assume a maximally spinning BH ($a=1$). Moreover, we set the logarithmic factor $\ln (R_{\rm out}/R_{\rm in})=1$ as the ejecta likely arises from the inner accretion flow, and neglect the difference between the poloidal and toroidal components of $B$ (as they are similar in the BP82 model). In this case, $\dot{M}_{\rm wind} < (B R_{\rm g})^2/c$, where $B = \phi (\dot{M}_{\rm acc} c)^{1/2}/(2 \pi R_g)$, and $\phi \approx 50$ for the most extreme case \citep{davis_tchekhovskoy_2020}. This gives $\dot{M}_{\rm wind} < (\phi/(2\pi))^{2} \dot{M}_{\rm acc}$ which for our case gives $\dot{M}_{\rm wind} \lesssim 4 \times 10^{20} \; {\rm g \, s^{-1}}$. This is a naive estimate, based on X-ray model-dependent mass accretion rate estimates, but it shows that the BP82 disk wind model is capable of launching the wind required by our modelling if the mechanism is efficient.

\par
The total ejection energy is $E \approx 3.5\times 10^{43}$ erg. The required jet power over the 5h launching duration is $L_{\rm jet} \approx 2\times 10^{39}$ erg s$^{-1}$. This is consistent with the maximum possible MAD jet power \citep{davis_tchekhovskoy_2020} for a BZ77 mechanism assuming a maximally spinning BH of $\dot M_{\rm acc} c^2 \approx 6\times 10^{39}$ erg s$^{-1}$. However, the BZ77 mechanism predicts the jet is initially Poynting-flux dominated and only later can be loaded by mass \citep{Riordan_2018}. Given the similarity of the derived values of $\dot{M}_{\rm ejecta}$ and $\dot{M}_{\rm acc}$, it remains a possibility that a large fraction of the accreted mass is entrained by a Blandford-Znajek jet.

\subsection{Reverse shock}
In this work we have shown that while the forward shock is required for late-time emission, the reverse shock dominates bright, early emission from ejecta, in agreement with analytic estimates \citep{matthews_2025} and relativistic hydrodynamic simulations \citep{savard_2025}. This is most easily demonstrated by how poorly the forward-shock-only Model (A3) performs (see Section \ref{sect:further_investigation}), primarily due to the steep decline of the early lightcurve. This finding is in agreement with conclusions of previous work \citep{wang_2003,hao_2009}, where it was concluded that a forward shock would evolve too slowly to explain the observed lightcurve evolution for the ejecta from XTE J1550-564. 

\subsubsection{Crossing timescale}
Generically, the reverse shock crossing time is pinpointed by the first peak of emission. In our best-fit lightcurves, all models typically infer a reverse shock crossing time slightly later than the first datapoint (MJD 58090.78), which was the brightest observed. 
This sharp rise during the reverse shock crossing explains why the ejecta was not detected on MJD 58059 and MJD 58080, where ATCA observations were taken with a sufficiently high angular resolution (the beamsize along the jet axis was around $1$ arcsec in both observations) to spatially resolve the ejecta from the core \citep[see further discussion in][ section 3.4]{russell_2019}. Nonetheless, earlier time observations, particularly covering the reverse shock crossing at lower frequencies, could be invaluable for future modelling. This is especially important in determining the initial Lorentz factor, ejecta energy, and ISM density which are the primary parameters setting the reverse shock crossing timescale. 
\par
The peak of the reverse shock emission coincides temporally with a significant sudden reduction in the ejecta separation measurement $\sim$90 days post launch. As the reverse shock crosses the ejecta, material at the front will radiate and cool before material at the back, resulting in the best-fit centroid may appear to slow or possible reverse direction. Furthermore, the transition between reverse and forward shock dominated emission zones may result in the opposite effect. Higher angular resolution observations around this crucial phase, (assuming emission is not resolved out), will help elucidate whether additional information regarding the blastwave geometry can be obtained through the observations of the reverse-forward shock transition. Full interpretation of such observations will require improved modelling through distinct modelling of the core-separation of reverse and forward shocked material, or synthetic radio images of ejecta simulations \citep{wang_2024,savard_2025}. 

\subsubsection{Efficiency}
Our best-fit Model (A) implies an unexpectedly low magnetization for the reverse shocked material, in stark contrast with GRB models which typically require $\epsilon_{\rm B, RS} > \epsilon_{\rm B, FS}$. The results of Model (A6) imply our underlying assumption of the geometry of the reverse shock is approximately correct, and cannot explain the low magnetization as a model-dependent effect. It is plausible this result may reflect a fundamental difference between reverse shocks in violent GRBs and relatively mild accretion energetics of XRB ejecta. However, further testing, especially the satisfactory fit to the data obtained by Model (A5), suggest that our finding could be merely a manifestation of a low value of $\epsilon_{\rm e, RS}$ (the fraction of energy transferred to the acceleration of electrons). We conclude that the reverse shock is likely characterised by either low magnetization or low particle acceleration efficiency, which may be tested by future modelling of other BH-XRB sources. 

\subsection{Under-dense interstellar medium}
In all models, we ubiquitously find interstellar media densities much lower than the canonical ISM density of $n \sim 1 {\rm cm}^{-3}$. This finding supports the long-standing hypothesis that the environment of XRBs, at least along the jet axis of sources which produce discrete ejecta, are characterised by an under-dense media \citep{heintz_underdense_2002,carotenuto_2024}. Moreover, there is evidence that the most powerful, furthest propagating ejecta may reach the edge of such a cavity \citep{hao_2009,carotenuto_2022a, zdziarski_2023}. Such cavities are suggested to have formed due to previous jet activity \citep{gallo_2005,carotenuto_2022a,sikora_2023,savard_2025}. There is tentative evidence that the supernova remnant (SNR) G323.7-1.0 was associated with the formation of MAXI J1535 and may have resulted in the underdense cavity \citep{maxted_2020}. The cavity's observed distance of $3.5$kpc and spatial size of tens of arcminutes appears to agree with our distance estimate and our assumption that ejecta-cavity interaction is not significant for this source. However, XRBs with low mass companions are generally expected to be relatively old (100s Myr), such that the SNR associated with BH formation is no longer observable; therefore, further evidence is required for a robust association.

\subsection{Jet profile and spreading}
Our model comparison shows clear evidence that the MAXI J1535 data is best fit by a jet which does not undergo significant lateral spreading. Inspection of best-fit kinematics plots (see Appendix \ref{app:plots}) imply the late-time data, particularly the final datapoints on MJD 58393, are consistently under predicted by models where spreading is enabled. Lateral jet spreading likely occurs due to transverse motion induced by internal pressure gradients acting on the surrounding media, and therefore is expected to become significant only after a reverse shock heats jet material. However, the reverse shock crossing time represents a much larger fraction of the overall evolution for XRB jets than GRBs \citep{matthews_2025}, owing in part due to the 
low sound speed $c_{\rm s} \ll \beta_{\rm jet}$. For this reason, jet spreading is possibly much less significant for mildly relativistic blast waves, including XRBs ejecta. Our findings also show significant, albeit less decisive, preference for tophat jets over Gaussian profile jets, although given the lower Bayes factor $\mathcal{B} = 223$, further evidence is required to confirm this.

\subsection{Inclination angle}
All models find a large jet viewing angle of $\theta_{\rm obs} = 65-80$ deg, in good agreement with the consensus from X-ray disc reflection features. This suggests that jet axis and the disc face are likely perpendicularly aligned. 

Our finding of a larger $\theta_{\rm obs}$ than previous works implies an initial Doppler factor for the approaching jet (assuming $\Gamma \lesssim 1.4$ and $\theta_{\rm obs} \gtrsim 60$ degrees) of $\delta_{\rm app} \lesssim 1.1$ and $\delta_{\rm rec} \gtrsim 0.5$. We can crudely estimate the peak receding flux density, taking into account our derived spectral index of $\alpha \approx 0.75$, using the fact that $f_{\rm rec} \approx f_{\rm app} \delta_{\rm rec}^{3-\alpha}/\delta_{\rm app}^{3-\alpha}$. The peak observed flux density of $f_{\rm app} = 2.87$ mJy corresponds to an estimated receding jet flux of $f_{\rm rec} \gtrsim 0.15$ mJy, depending on the deceleration profile and real value of $\theta_{\rm obs}$. In the case of MAXI J1535, no receding jet was detected in the observations. Such a detection may be confounded by the receding jet having a smaller core-separation, greatly exacerbated by the poor angular resolution of early ATCA observations \citep{russell_2019}, and thus the receding ejecta component may be difficult to distinguish from core emission which remained bright until MJD 58150. Future modelling may be improved by incorporating non-detections of receding jets, particularly with high resolution instrumentation, to further constrain the jet inclination angle and initial Lorentz factor. A detailed discussion on observing receding jets will be presented in a future work.

\subsection{Caveats}
\label{sect:caveats}
There are a number of caveats to this work related to the limitations of the model. The first major caveat is our assumption that the reverse shock is co-spatial with the forward shock in volume and core-separation. This is required in the model such that an emissivity can be calculated in the \texttt{jetsimpy} custom radiation model, and is generally motivated by the fact that the outflow width is much less than the ejecta propagation distance. Nevertheless, we interrogate this assumption in our Model (A6) where the shock width (and effectively, the volume of reverse-shocked material) is a free parameter, and find this likely plays a minor role. 
\par
In the core-separation modelling, we use the \texttt{jet.Offset} function in \texttt{jetsimpy}, which strictly tracks only the core-separation of the forward shock emission region. However, this is a valid approximation during and shortly after the reverse shock crossing, as the forward and reverse shocks have very similar core-separations (see Fig. 9 in \citealt{matthews_2025}). Crucially, this is precisely when we expect the reverse shock emission dominates over the forward shock, and thus we do not expect this to significantly affect our results. However, an exciting prospect is that future high resolution observations could observed distinct components corresponding to the forward and reverse shock, particularly in the timeframe directly after the crossing where both shocks significant contribution to emission. 
\par
Another caveat to this work is the assumed uniform density profile of the ISM. \cite{carotenuto_2022a} found that a two-stage cavity/ISM media best-fit the kinematic data of MAXI J1348-630. Additional deceleration upon jet-cavity interaction may result in late-time rebrightenings as observed from MAXI J1535 around 380 days post-launch. Although our findings clearly require reverse shock emission due lack of resolved jet ejecta before 85 days post-burst and the sharp decay of the lightcurve (see discussion in \ref{sect:further_investigation}, also \citealt{wang_2003}), we consider it concievable, but very unlikely, that the reverse shocked material could persist to produce observed late-time emission due to jet-ISM interactions. This is disfavoured as late-time rebrightenings likely require in-situ particle acceleration \citep{bright_2020}, which does not occur within the actual ejecta material after the reverse shock has already crossed. Furthermore, core-separation of reverse-shocked material slows drastically at late times \citep{matthews_2025} making it difficult to reach observed late-time separation. Nonetheless, future joint radiative and kinematic modelling may include models containing a cavity via changes to the ISM density profile, to verify our conclusion that both shocks are required to explain the data. Finally, a simple way to verify the presence of forward shock emission at late-times would be to conduct a single deep observations at late-times, after non-detection in shorter integration time observations. Jet-cavity interactions of reverse shocked material would fade much faster than a forward shock, which evolves on longer timescales. 
\par
The final caveat to this work is that the derived distance is found to be at the lowest end of the prior distribution in all models. The prior is enforced based on independent and robust observational constraints obtained by \cite{chauhan_2019}, and therefore is likely representative of a systematic in the model. If indeed MAXI J1535 is located at a distance closer than $3.5\, {\rm kpc}$, the observed luminosity and core-jet separation will increase, and therefore will result in a lower value of $E_0$ (and commensurately higher value of $n_{\rm ism}$).

\section{Conclusions \& Outlook}
\label{sect:conclusion}
In this work, we have performed joint radiative and kinematic modelling of XRB ejecta from MAXI J1535. We successfully fit both the flux and kinematic data from this source, demonstrating how combining flux and separation data for these sources allows us to break key energetic degeneracies, significantly improving parameter estimation. Our primary conclusions are as follows:
\begin{enumerate}
    \item The reverse shock dominates the MAXI J1535 ejecta emission at early times and is characterised by either low magnetisation or inefficient particle acceleration.
    \item The jet has a moderate initial Lorentz factor $\Gamma \approx 1.4$, does not undergo significant lateral spreading, and is likely launched perpendicular to the disc.
    \item The initial ejecta energy is $E_{0} \gtrsim {\rm few} \times 10^{43}$ erg and may be powered by the accretion disc alone.
    \item The ejecta propagates into a low $n_{\rm ism} < 10^{-4}$ cm$^{-3}$ density environment.
\end{enumerate}

These findings are in good agreement with previous work \citep{wang_2003,carotenuto_2024}, and motivate future theoretical and observational investigation. Modelling of additional XRB ejecta sources with sufficient temporal and frequency coverage will be essential to ascertain whether the findings of this work apply ubiquitously. Sources with early-time observations covering the rise corresponding to the reverse shock crossing, those with high luminosities that may be incompatible with disc-powered jets, and sources where a receding jets are also detected represent the most interesting sources to which similar modelling could be applied. Future work would benefit from the development of dedicated blast-wave models which produce kinematic and radiative predictions for trans-relativistic, off-axis jets which explicitly include the reverse shock. This could be realised through the development of an MHD simulation grid from which radiation and kinematic predictions could be fit to the data by interpolating between models. Crucially, such an approach could take advantage of kinematic degeneracies (e.g., between $E_{0}$ and $n_{\rm ism}$) to reduce the total number of runs required to cover the parameter space. Finally, polarization measurements throughout the ejecta evolution may represent an additional dimension for modelling, if robust theoretical predictions can be folded into the model.

\section*{Acknowledgements}
The authors would like to thank the anonymous referee for useful comments which improved this manuscript. AJC thanks H. Wang for the publication of, and advice in using, \texttt{jetsimpy}, and H. Gao for useful insights regarding reverse shock models. AJC also acknowledges fruitful discussions with D. Aksulu, A. van der Horst, A. Hughes, and the XKAT Collaboration. 
\par
AJC acknowledges support from the Oxford Hintze Centre for Astrophysical Surveys which is funded through generous support from the Hintze Family Charitable Foundation. JHM acknowledges funding from a Royal Society University Research Fellowship (URF\textbackslash R1\textbackslash221062). RF acknowledges support from UK Research and Innovation, The European Research Council, and the Hintze Family Charitable Foundation. GPL is supported by a Royal Society Dorothy Hodgkin Fellowship (grant Nos. DHF-R1-221175 and DHF-ERE-221005). TDR acknowledges support as an INAF IAF research fellow. NS acknowledges support from the Knut and Alice Wallenberg Foundation through the ``Gravity Meets Light'' project and by the research environment grant ``Gravitational Radiation and Electromagnetic Astrophysical Transients'' (GREAT) funded by the Swedish Research Council (VR) under Dnr 2016-06012. KS acknowledges support from the Clarendon Scholarship Program at the University of Oxford and the Lester B. Pearson Studentship at St John's College, Oxford. AAZ acknowledges support from the Polish National Science Center grants 2019/35/B/ST9/03944 and 2023/48/Q/ST9/00138.

\section*{Data Availability}
All observational data presented in this work are freely available in the referenced material, but will also be provided upon request to the authors. A full reproduction package, including theoretical models, data products, and plots will be made available at: \url{https://github.com/alexanderjsc} after publication.



\bibliographystyle{mnras}
\bibliography{example} 



\appendix

\section{Constraining the jet inclination angle for a given proper motion}
\label{app:inclination}
\begin{figure}
\begin{center}
\includegraphics[width=0.5\textwidth]{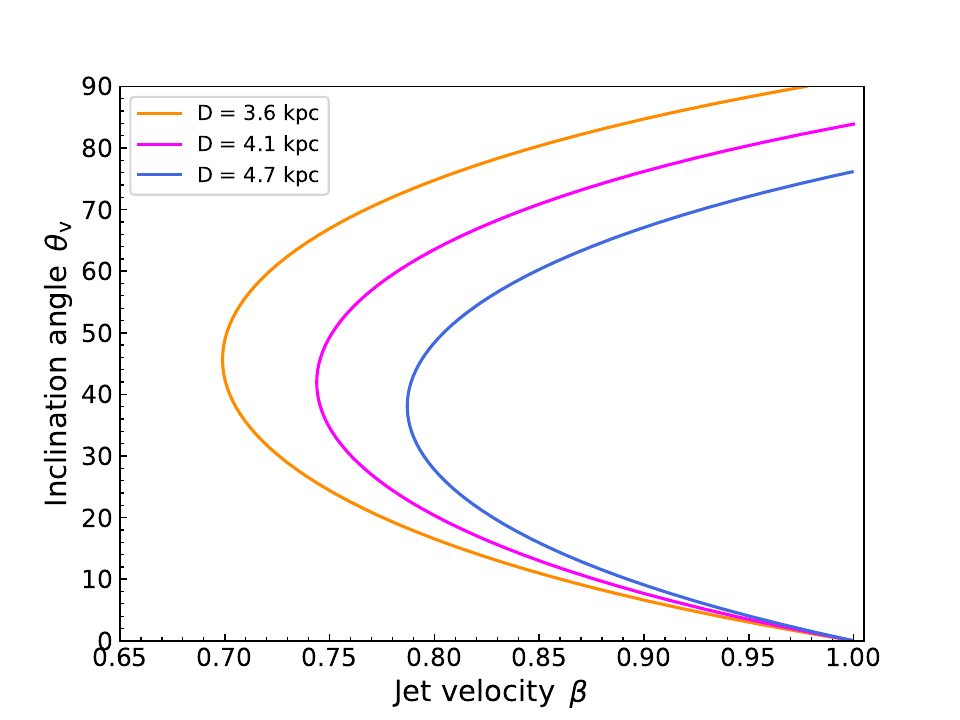}
\caption{Updated inclination angle as a function of jet velocity $\beta = v/c$ for the maximum observer proper motion from MAXI J1535 of 47 mas/day.}
\label{fig:app:inclination_angle_new}
\end{center}
\end{figure}

As discussed in subsection \ref{sect:inclination_angle_prediscussion}, we re-examined constraints on the jet angle to the line of sight ($\theta_{\rm obs}$) presented in \cite{russell_2019}, and found additional solutions which allowed a larger range of $\theta_{\rm obs}$. In Fig. \ref{fig:app:inclination_angle_new} we present our new $\theta_{\rm obs}$-$\beta$ solutions, recreating fig. 9 of \cite{russell_2019}. 

The new solutions exist above an inflection point, representing an upper branch for which similar values of $\beta$ can result in larger inclination angles. Generically, one can derive a threshold value of proper motion as a function of source distance of 173.5 mas day$^{-1}$ kpc$^{-1}$, above which the inclination angle can be constrained to $\theta_{\rm obs} < 90$ degrees (see also \citealt{zdziarski_2025b}). This is equivalent to the condition for apparent superluminal motion. We include the updated constraints for the assumed distances in Table \ref{tab:new_inc_constraints} and also update the inclination angle constraints for MAXI J1348-630 from \cite{carotenuto_2021} for which similar extension to the published solutions exists. Finally, we note that in the kinematic-only modelling of \cite{carotenuto_2024}, the authors best-fit inclination angle $\theta_{\rm obs} = 30.3\pm6.3$ deg is much lower than the typical values in this work. However, this is not in tension as it appears, due to the degeneracy about a critical inflection point in Fig. \ref{fig:app:inclination_angle_new}. The derived value of $\beta$ is similar for both models, as would be expected given the similar kinematic profile.

\setlength{\tabcolsep}{5pt}
\setlength{\extrarowheight}{.9em}
\begin{table}
\centering
\begin{tabular}{l|l|l}
\hline
\hline
\textbf{Source}  &\textbf{Distance} [kpc] & \textbf{Constraint} [deg] \\ \hline 
MAXI J1535-571 & 3.6 & n/a \\ \cline{2-3} 
                           & 4.1 & 83.3  \\   \cline{2-3} 
                           & 4.6  & 75.6  \\ \hline
MAXI J1348-630 & 1.6  & 89.6 \\ \cline{2-3} 
                                & 2.2 & 71.4  \\   \cline{2-3} 
                                & 2.7 & 60.5  \\ \hline
\end{tabular}
\caption{Updated jet angle to the observer line of sight proper motion upper limit constraints assuming the maximum best-fit proper motion values of 47 mas/day (MAXI J1535-571; \citealt{russell_2019}) and 108 mas/day (MAXI J1348-630; \citealt{carotenuto_2021}).}
\label{tab:new_inc_constraints}
\end{table}

\section{Hierarchy of Critical Frequencies}
\label{app:criticalfreqs}
\begin{figure}
\begin{center}
\includegraphics[width=0.5\textwidth]{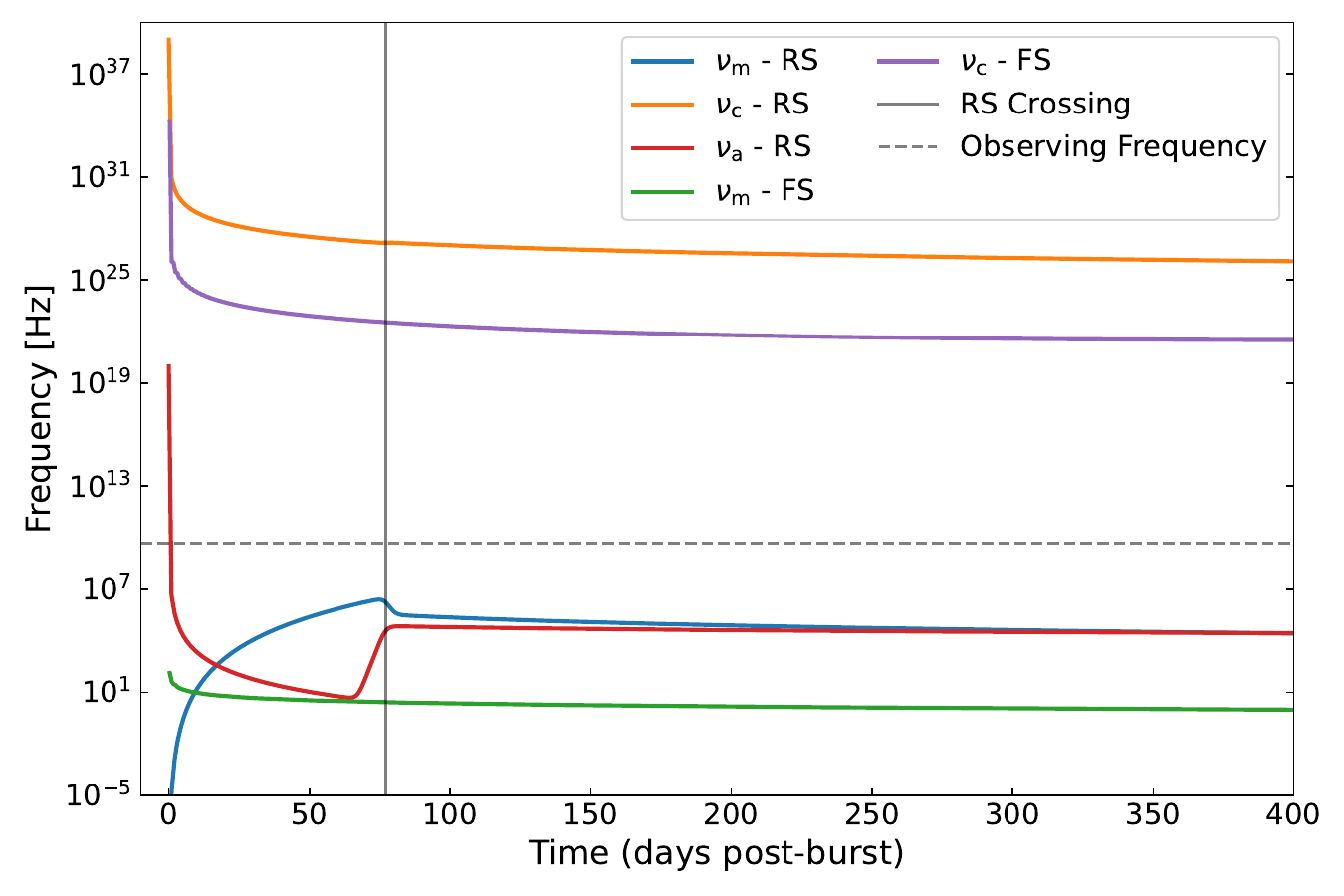}
\caption{Evolution of the peak, self-absorption, and cooling synchrotron frequencies for the preferred Model (A). The critical frequencies are are averaged in 800 second bins to account for different values at different regions in the hydrodynamic shock front, and for the reverse shock interpolated across the shock crossing time for readability.}
\label{fig:critical_frequencies}
\end{center}
\end{figure}

In order to model the lightcurve of the ejecta using \texttt{jetsimpy}, the critical synchrotron frequencies (absorption $\nu_{\rm a}$, minimum/peak $\nu_{\rm m}$, and cooling $\nu_{\rm c}$) are computed separately for the forward and reverse shock as a function of the time-dependent blastwave properties. For the forward shock we use the standard \texttt{sync\_dn} model in \texttt{jetsimpy} \citep{wang_2024}, and for the reverse shock we employ the prescription developed by \citet{kobayashi_2000a} as presented in \citet{gao_2013}. In Fig. \ref{fig:critical_frequencies} we show the evolution of all relevant critical frequencies as a function of observer-frame time for the preferred Model (A). The forward shock model in \texttt{jetsimpy} does not account for synchrotron self-absorption, so $\nu_{\rm a, FS}$ is not included in the plot. However in all models the forward shock is not relevant until late-times, where it is characterised by lower flux densities and larger spatial scales than the reverse shock, meaning that we can safely assume that $\nu_{\rm a, FS} < \nu_{\rm a, RS} < \nu_{\rm obs}$.

\section{Additional plots}
\label{app:plots}
\begin{figure*}
\begin{center}
\includegraphics[width=1.0\textwidth]{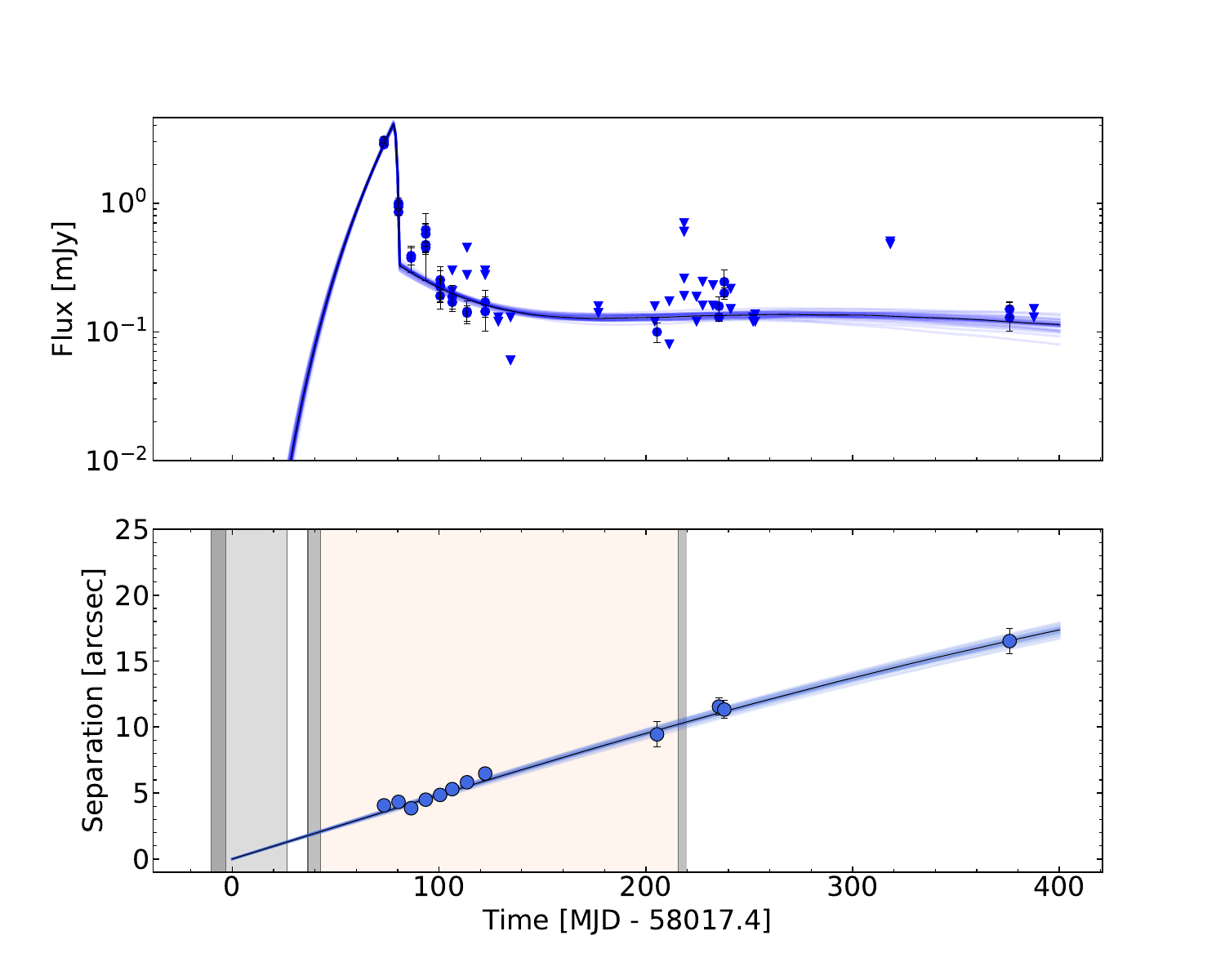}
\caption{Model (A) lightcurve (data and model scaled to $5.5$ GHz assuming $\alpha = 0.75$), and kinematics obtained from a sample of the posterior parameter distribution.}
\label{fig:multi_lc_model_a}
\end{center}
\end{figure*}

All best-fit lightcurve and kinematic plots from all tested models will be made available online after publication. In this Appendix we include a few useful diagnostic plots. In Fig. \ref{fig:multi_lc_model_a} we show a random sample of 30 lightcurves and separations from Model (A) obtained from the posterior parameter distribution, to give an indication of model uncertainties. For clarity, fluxes and upper limits in the top panel are scaled to a common frequency of $5.5$ GHz. The reverse shock crossing time is pinpointed by the available data, but some discrepancies are observed in the posterior distribution at late-times, likely owing to the lack of data and the uncertainties associated with faint detections. 

\begin{figure*}
\begin{center}
\includegraphics[width=1.0\textwidth]{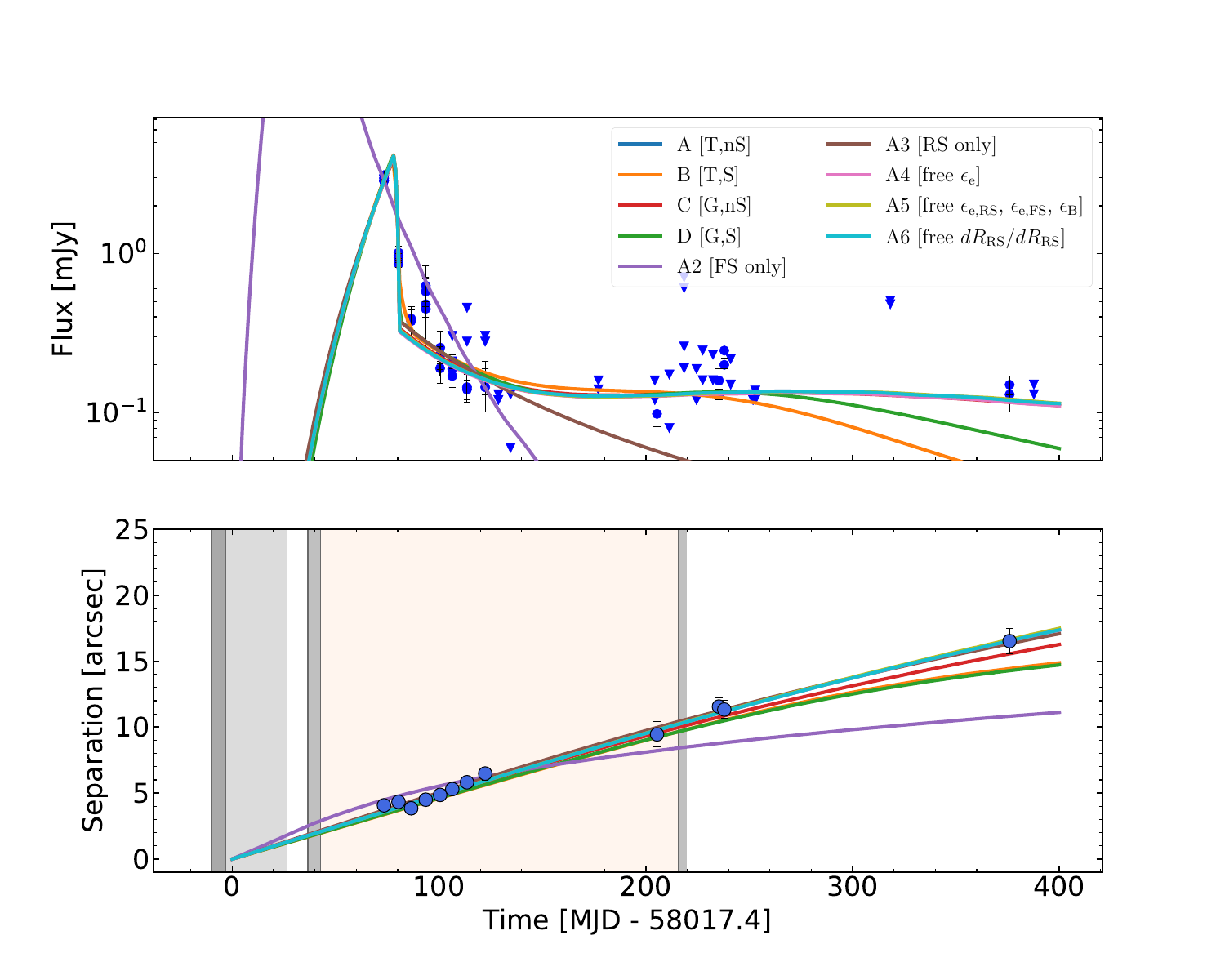}
\caption{Best-fit lightcurves (data scaled to $5.5$ GHz assuming $\alpha = 0.75$, irregardless of the best fit $p$ for individual models) and kinematics for all models. }
\label{fig:all_best_fit_curves}
\end{center}
\end{figure*}

In Fig. \ref{fig:all_best_fit_curves} we show the best fit lightcurves and separations for all models. With the notable exception of the FS-only (A3) and RS-only (A2) models, all follow similar lightcurve patterns with the reverse shock peaking at the crossing time at 90 days post-burst, and a late-time rebrightening powered by the forward shock. The forward shock only Model (A3) predicts a very bright early lightcurve, fails to explain late-time detections, and does not propagate to the required separation and late-times as discussed in Section \ref{sect:further_investigation}. The reverse shock only model does much better, but also fails to capture the late-time re-brightening. In general, the two spreading jet models (B) and (D) struggle to reproduce the final datapoints for both the lightcurve and kinematic profiles. All other models (A, A4, A5, A6, C) are non-spreading jets with different free parameters setting microphysical radiation, are somewhat comparable. Finally, we note the lightcurve for Model (A5) is plotted from 60 days and onwards as it is (exceptionally) self-absorbed until this time, likely due to the much larger $\epsilon_{\rm B}$, for which the flux is not well-defined in the model.

\begin{figure*}
\begin{center}
\includegraphics[width=\textwidth]{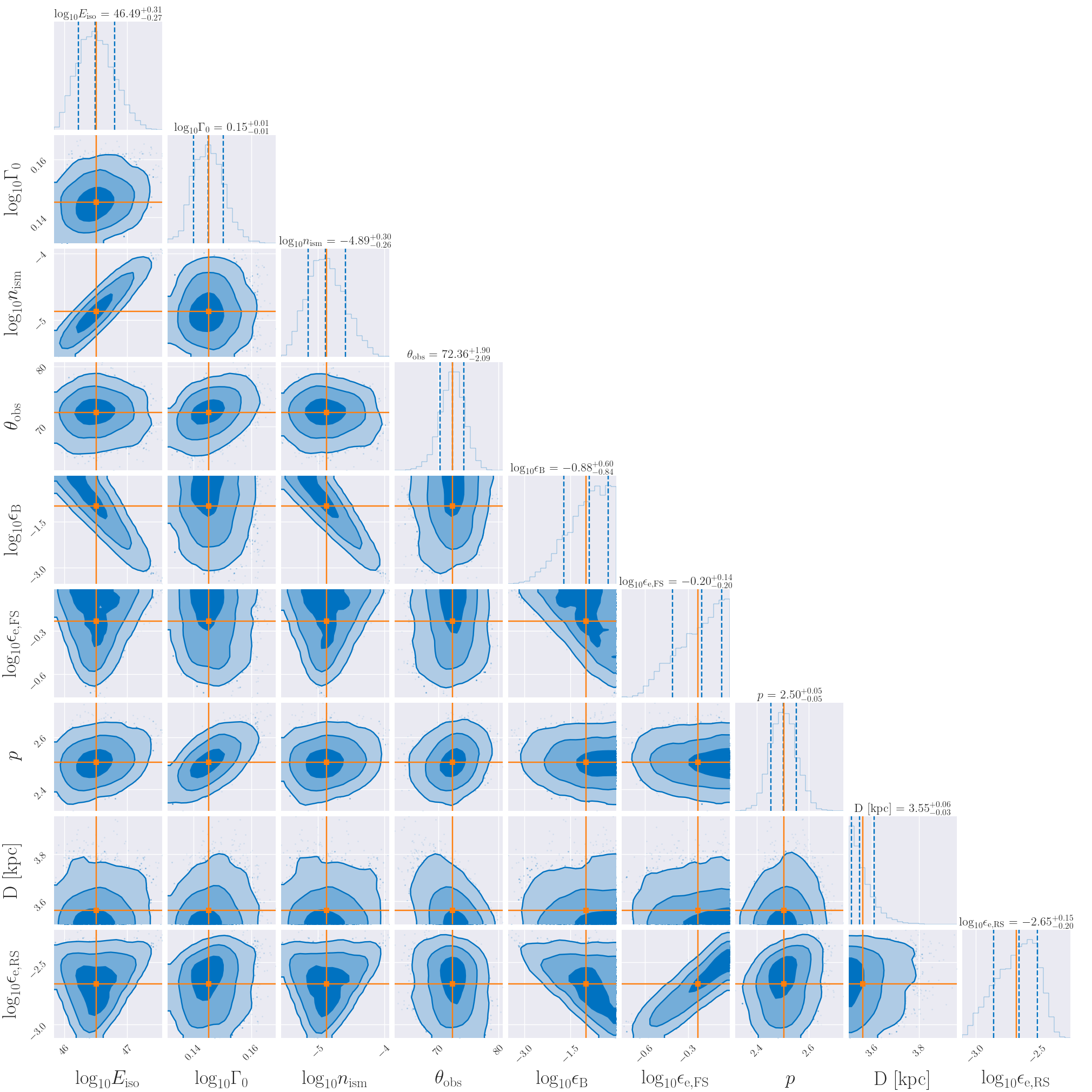}
\caption{Corner plot of the posterior distributions for Model (A5), the model which had the best reduced $\chi^2$.}
\label{fig:corner_A5}
\end{center}
\end{figure*}

Finally, in Fig. \ref{fig:corner_A5} we show the corner plot for Model (A5), which produces the best $\chi^2$ value and the second best BIC score. This model finds relatively high value for $\epsilon_{\rm B} \sim 0.1$, but a much lower value for $\epsilon_{\rm e, RS} \sim 10^{-2.65}$, meaning we cannot conclude the reverse shock has low magnetization, as the data is almost equally well explained by a low value of $\epsilon_{\rm e, RS}$. 

\bsp	
\label{lastpage}
\end{document}